\documentclass[aps,prl,reprint,amssymb]{revtex4-1}

\usepackage{bm}
\usepackage{graphicx}
\usepackage{amsmath}
\usepackage{amsthm}
\usepackage{eucal}
\usepackage{amssymb}
\usepackage{mathrsfs}
\usepackage{enumerate}
\usepackage{xcolor}

\newcommand{\comments}[1]{}

\begin{document}

\title{Models of genetic drift as limiting forms of the Lotka-Volterra competition model}

\author{George W.~A.~Constable and Alan J.~McKane}

\affiliation{Theoretical Physics Division, School of Physics and Astronomy,
The University of Manchester, Manchester M13 9PL, United Kingdom}

\begin{abstract}

The relationship between the Moran model and stochastic Lotka-Volterra competition (SLVC) model is explored via timescale separation arguments. For neutral systems the two are found to be equivalent at long times. For systems with selective pressure, their behavior differs. It is argued that the SLVC is preferable to the Moran model since in the SLVC population size is regulated by competition, rather than arbitrarily fixed as in the Moran model. As a consequence, ambiguities found in the Moran model associated with the introduction of more complex processes, such as selection, are avoided.

\end{abstract}
\maketitle

The modeling of genetic drift --- the mechanism by which the genetic makeup of a population can change due to random fluctuations --- is frequently viewed in a different way to the other key genetic processes, such as mutation, migration and selection, since it requires an inherently stochastic approach. Although genetic drift was first illustrated using the Wright-Fisher model~\cite{wright_1931,fisher_1930}, many authors now use the more tractable Moran model~\cite{moran_1957}. This has the same long-time behavior as the Wright-Fisher model~\cite{blythe_mckane_models_2007}, and the same assumption that the size of the population, $N$, is fixed. This assumption, which serves as a proxy for the biological processes not included in these models which control the population size, is therefore made for both historical reasons as well as mathematical ones. The artificial nature of this assumption is of course recognized, for example Moran in his book prefaced the description of his model with a discussion of its possible relation to more realistic situations~\cite{moran_1962}. Others introduce an effective population size to replace variations in population size, for instance, by an average~\cite{halliburton_2004,li_c_c_1955}. Yet in these cases, the effect is still to retain the fixed size of the population. 

Certainly the assumption that $N$ is fixed is hardly ever questioned in the
many papers describing the extensive use to which the Moran model has been put in the last decade or so (see, for instance,~\cite{nowak_2004,traulsen_2005,muirhead_2009}). However the assumption results in a single rate parameter encompassing birth and death, making it impossible to tease apart effects resulting from the various processes. The ambiguities inherent in this approach are particularly apparent when trying to include the effects of selection in the model. 

In this Letter we will advocate a different starting point which allows us to address some of these questions. We adopt a more ecologically-oriented approach and begin from a population of $n_1$ haploid individuals which carry allele $A_1$ and $n_2$ haploid individuals which carry allele $A_2$. They will reproduce at rates $b_1$ and $b_2$ respectively and die at rates $d_1$ and $d_2$. We will also allow for competition between individuals of type $A_i$ and $A_j$, at a rate $c_{ij}$. This will tend to regulate the population size, without imposing the condition $n_{1}+n_{2}=N$. The model will be formulated as an individual based model (IBM), since the stochastic aspects are central to the discussion. 

Although it is quite easy to sketch out this idea, showing how precisely this model relates to conventional models in population genetics, such as the Moran model, is not so straightforward, and we are aware of only a few studies which touch on this issue. In \cite{noble_hastings_2011}, an exact mapping between the two models was sought via the introduction of unconventional fitness weightings, while \cite{parsons_quince_2007_b,parsons_quince_2008,parsons_quince_2010} were concerned with significant deviations from Moran phenomenology. To provide a systematic understanding of the relationship between the two approaches, we will apply an approximation procedure which we recently developed based on the elimination of fast modes~\cite{constable_phys,constable_bio}.

\begin{figure}
\setlength{\abovecaptionskip}{-2pt plus 3pt minus 2pt}
\begin{center}
\includegraphics[width=0.4\textwidth]{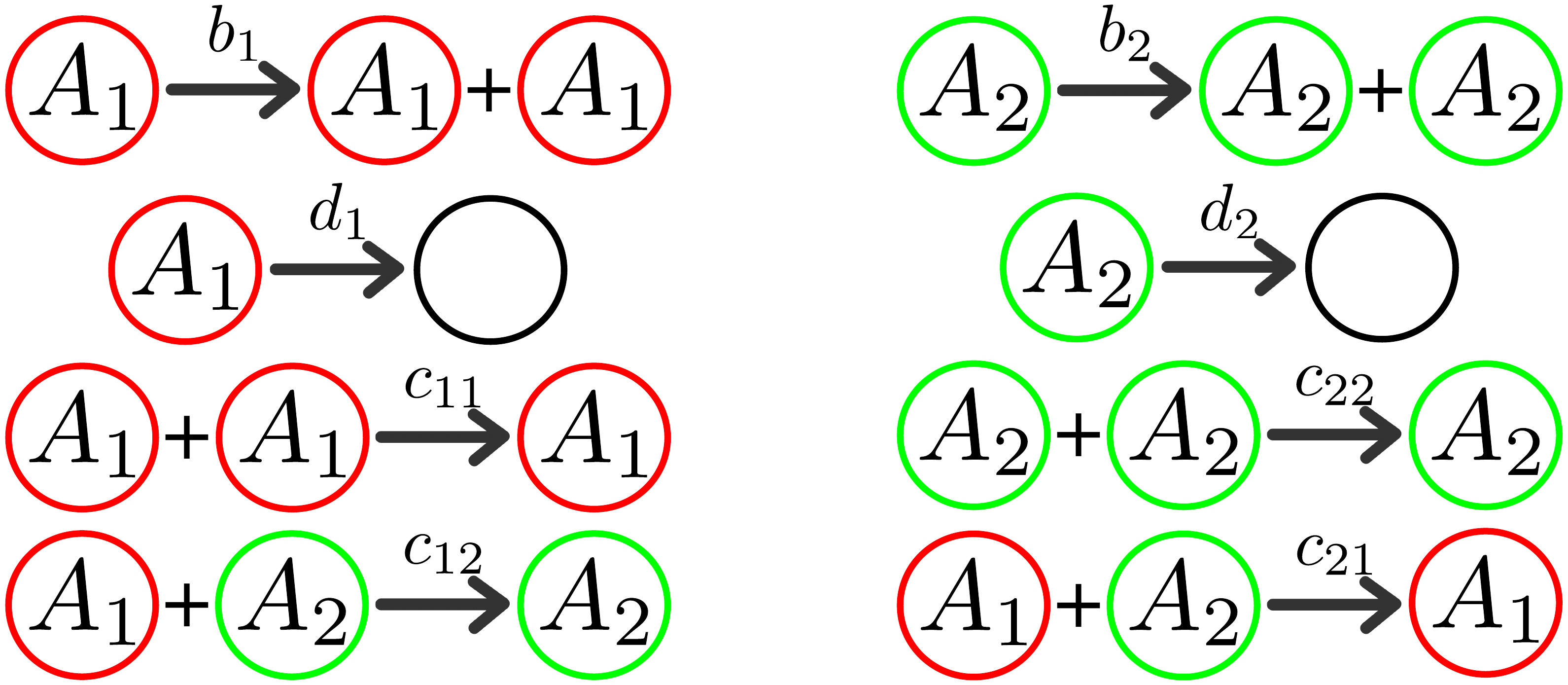}
\end{center}
\caption{Reactions specifying the SLVC model.}\label{fig_reactions}
\end{figure}

The system we will investigate will be well-mixed, and so its state will be completely specified by $\bm{n}=(n_1,n_2)$. This state will be able to change because of births, deaths or competition between individuals, defined by the rules given in Fig.~\ref{fig_reactions}. To define a dynamics we need to specify the rates at which the allowed changes in Fig.~\ref{fig_reactions} take place. We assume mass action for the competitive interactions leading to transition rates given by
\begin{eqnarray}\label{trans_rates}
T_1(n_1 + 1,n_2|n_1,n_2) &=& b_1\frac{n_1}{V}, \nonumber \\
T_2(n_1,n_2 + 1|n_1,n_2) &=& b_2\frac{n_2}{V}, \\
T_3(n_1 - 1,n_2|n_1,n_2) &=& d_1\frac{n_1}{V} + c_{11}\frac{n_1}{V}\frac{n_1}{V} 
+ c_{12}\frac{n_2}{V}\frac{n_1}{V}, \nonumber \\ 
T_4(n_1,n_2 - 1|n_1,n_2) &=& d_2\frac{n_2}{V} + c_{22}\frac{n_2}{V}\frac{n_2}{V} 
+ c_{21}\frac{n_1}{V}\frac{n_2}{V}. \nonumber
\end{eqnarray}
The parameter $V$ is not the total number of individuals in the system, which is free to vary. Rather it is a measure of the size of the system. Typically it would be an area or a volume, but its precise value or even its dimensions can be left unspecified, as they can be absorbed into the rates $b_i,d_i$ and $c_{ij}$. The probability of finding the system in state $\bm{n}$ at time $t$, $P_{\boldsymbol{n}}(t)$, may be found from the master equation~\cite{van_Kampen_2007}
\begin{eqnarray}
\frac{\mathrm{d}P_{\boldsymbol{n}}(t)}{\mathrm{d}t} = \sum^{4}_{\mu=1}
\left[ T_{\mu}(\boldsymbol{n}|\boldsymbol{n}-\boldsymbol{\nu}_{\mu})
P_{\boldsymbol{n}-\boldsymbol{\nu}_{\mu}}(t) - \right. \nonumber \\ \left.
T_{\mu}(\boldsymbol{n}+\boldsymbol{\nu}_{\mu}|\boldsymbol{n})
P_{\boldsymbol{n}}(t) \right],
\label{master_alt}
\end{eqnarray}
where $\boldsymbol{\nu}_{\mu}$ describes how many individuals of one type are transformed during the reaction $\mu=1,\hdots,4$. So, $\boldsymbol{\nu}_{1} = (1,0), \boldsymbol{\nu}_{2} = (0,1), \boldsymbol{\nu}_{3} = (-1,0)$ and $\boldsymbol{\nu}_{4} = (0,-1)$. Equations \eqref{trans_rates} and \eqref{master_alt}, together with an initial condition for $P_{\boldsymbol{n}}$, allow us in principle to find $P_{\boldsymbol{n}}(t)$ for all $t$. 

In practice, the master equation is intractable. To make progress the diffusion approximation is made, that is $V$ is assumed sufficiently large that $x_i \equiv n_i/V$ is approximately continuous~\cite{crow_kimura_into}. We can then expand the master equation as a power series in $V^{-1}$ to obtain the Fokker-Planck equation (FPE)~\cite{gardiner_2009}
\begin{eqnarray}
\frac{\partial P(\bm{x},\tau)}{\partial \tau} &=& 
- \sum_{i=1}^2 \frac{\partial }{\partial x_i} 
\left[ A_{i}(\bm{x}) P(\bm{x},\tau) \right] \nonumber \\
&+& \frac{1}{2V} \sum_{i,j=1}^2 \frac{\partial^2 }{\partial x_i \partial x_j} 
\left[ B_{i j}(\bm{x}) P(\bm{x},\tau) \right], 
\label{FPE}
\end{eqnarray}
where $\tau = t/V$ is a rescaled time and where we have neglected higher order terms in $V^{-1}$. The functions $A_i$ and $B_{ij}$ can be expressed in terms of the $\nu_{i,\mu}$ and functions $f_\mu$ as~\cite{mckane_BMB}
\begin{equation}
A_i(\bm{x}) = \sum_{\mu=1}^4 \nu_{i,\mu}f_{\mu}(\bm{x}), \ \
B_{i j}(\bm{x}) =\sum^{4}_{\mu=1} \nu_{i,\mu} \nu_{j,\mu}
f_{\mu}(\bm{x}),
\label{A_and_B}
\end{equation}
where $i,j=1,2$ and where the functions $f_{\mu}(\bm{x})$ are equal to $T_{\mu}(V\boldsymbol{x}+\boldsymbol{\nu}_{\mu}|V\boldsymbol{x})$. The diffusion approximation, made popular by Kimura and others in the context of population genetics~\cite{crow_kimura_into}, is usually expressed in  the form of an FPE such as Eq.~\eqref{FPE}, however for our purposes it is preferable to work with the entirely equivalent It\={o} stochastic differential equations (SDEs)~\cite{gardiner_2009}
\begin{equation}
\frac{\mathrm{d}x_i}{\mathrm{d}\tau} = A_i(\bm{x}) + \frac{1}{\sqrt{V}} \eta_i(\tau),
\label{SDE_orig}
\end{equation}
where $\eta_{i}(\tau)$ is a Gaussian noise with 
\begin{equation}
\left\langle \eta_i(\tau)\right\rangle =0 \,, \qquad \, \left\langle \eta_i(\tau) \eta_j(\tau') \right\rangle = B_{i j}(\bm{x}) \delta\left( \tau - \tau' \right).
\label{correlator}
\end{equation}
\begin{figure}
\setlength{\abovecaptionskip}{-2pt plus 3pt minus 2pt}
\begin{center}
\includegraphics[width=0.4\textwidth]{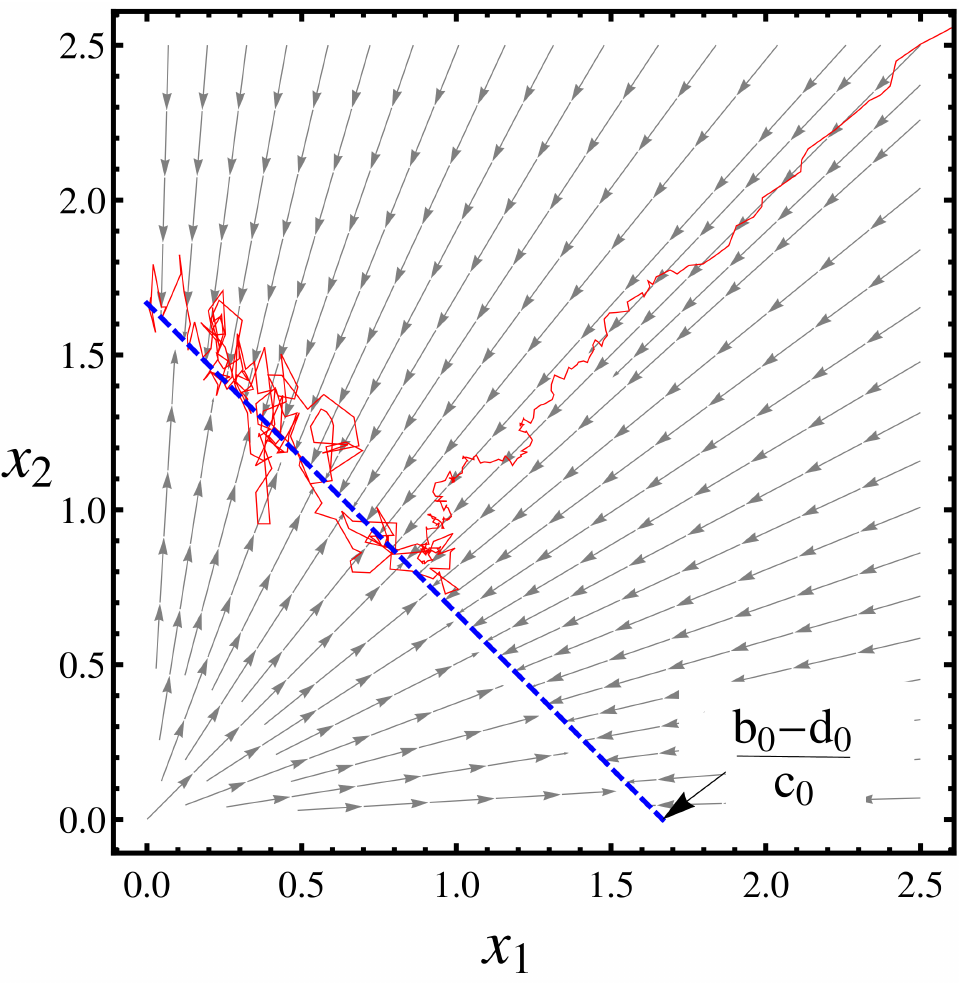}
\end{center}
\caption{(Color online) A stochastic simulation of the model specified in Fig.~\ref{fig_reactions} is plotted in red, along with the mean deterministic behavior (given by $A_{i}(\bm{x})$ in Eqs.~\eqref{explicitA_B}) in gray. Parameters used here are for the neutral model, with $b_{0}=2$, $d_{0}=1$, $c_{0}=0.6$ and $V=300$. The stochastic system follows an approximately deterministic trajectory until it reaches the center manifold (CM), plotted in blue and given by Eq.~\eqref{centre_manifold}.}\label{fig_traj_neutral}
\end{figure}
The precise form of the functions $A_i(\bm{x})$ and $B_{i j}(\bm{x})$ for the system of interest to us can be read off from Eq.~\eqref{A_and_B} using Eq.~\eqref{trans_rates} and the $\nu_{i,\mu}$ given earlier. One finds
\begin{eqnarray}
A_1(\bm{x}) &=& \left( b_1 - d_1 \right)x_1 - c_{11} x^2_1 - c_{12} x_1 x_2 \,, 
\nonumber \\
A_2(\bm{x}) &=& \left( b_2 - d_2 \right)x_2 - c_{21} x_1 x_2 - c_{22} x^2_2 \,, 
\nonumber \\
B_{11}(\bm{x}) &=& \left( b_1 + d_1 \right)x_1 + c_{11} x^2_1 + c_{12} x_1 x_2 
\,, \nonumber \\
B_{22}(\bm{x}) &=& \left( b_2 + d_2 \right)x_2 + c_{21} x_1 x_2 + c_{22} x^2_2 
\,, 
\label{explicitA_B}
\end{eqnarray}
and $B_{ij}=0$, for $i \neq j$. In the limit $V \to \infty$, Eq.~\eqref{SDE_orig} reduces to the two deterministic differential equations $\mathrm{d}x_i/\mathrm{d}\tau = A_i(\bm{x})$, with $A_i(\bm{x})$ given by Eq.~\eqref{explicitA_B}, which are the familiar Lotka-Volterra equations for two competing species~\cite{roughgarden_1979,pielou_1977}.

We begin the analysis by assuming that individuals of type $A_1$ and $A_2$ have equal fitness. Thus the theory is neutral, and $A_1$ and $A_2$ have equal birth, death and competition rates: $b_i \equiv b_0, d_i \equiv d_0, c_{ij} \equiv c_0$. Simulations of the original IBM defined by Fig.~\ref{fig_reactions} and Eq.\eqref{trans_rates} are shown in Fig.~\ref{fig_traj_neutral}, where it is seen that the trajectories quickly collapse onto a line in the $x_1$-$x_2$ plane. This can be understood by first considering the deterministic trajectories (shown in gray in Fig.~\ref{fig_traj_neutral}). We begin by looking for fixed points of the dynamics by setting $A_i(\bm{x})=0, i=1,2$. Taking the combinations $A_1 \pm A_2$ we find that the fixed points are solutions of the two equations
\begin{equation}
\left[ \left( b_0 - d_0 \right) - c_0\left( x_1 + x_2 \right) \right] 
\left( x_1 \pm x_2 \right) = 0.
\label{fixed_points}
\end{equation}
We see that, apart from the trivial fixed point $x_1 = x_2 = 0$, there is a line of fixed points given by
\begin{equation}
x_1 + x_2 = (b_0 - d_0)c_0^{-1}.
\label{centre_manifold}
\end{equation}
This is the equation of the blue line shown in Fig.~\ref{fig_traj_neutral}. Further insight can be gained by calculating the Jacobian at points on Eq.~\eqref{centre_manifold}. One finds that it has eigenvalues $\lambda^{(1)}=0$ and $\lambda^{(2)}=-(b_0 - d_0)$, with corresponding eigenvectors 
\begin{align}\label{eigenvectors_1}
\bm{u}^{(1)} = \frac{c_0}{(b_0 - d_0)}\,\left( \begin{array}{c} x_2 \\ - x_1 \end{array} \right), \ \ \bm{v}^{(1)} = \left( \begin{array}{c} 1 \\ - 1 \end{array} \right),
\end{align}
and 
\begin{align}\label{eigenvectors_2}
\bm{u}^{(2)} = \left( \begin{array}{c} 1 \\ 1 \end{array} \right), \ \ 
\bm{v}^{(2)} = \frac{c_0}{(b_0 - d_0)}\,\left( \begin{array}{c} x_1 \\ x_2 \end{array} \right),
\end{align}  
where $\bm{u}^{(i)}$ and $\bm{v}^{(i)}$ are respectively the left- and right-eigenvectors corresponding to the eigenvalue $\lambda^{(i)}$ (normalised such that $\sum^{2}_{k=1}\,u^{(i)}_k v^{(j)}_k = \delta_{i j}$), and $x_{2}$ is given by Eq.~\eqref{centre_manifold}. This shows that Eq.~\eqref{centre_manifold} defines a CM to which the deterministic system quickly collapses; there is then no further motion along this line. The timescale for the collapse of this fast mode is given by $| \lambda^{(2)}|^{-1} = (b_0 - d_0)^{-1}$. 
\begin{figure}
\setlength{\abovecaptionskip}{-2pt plus 3pt minus 2pt}
\begin{center}
\includegraphics[width=0.4\textwidth]{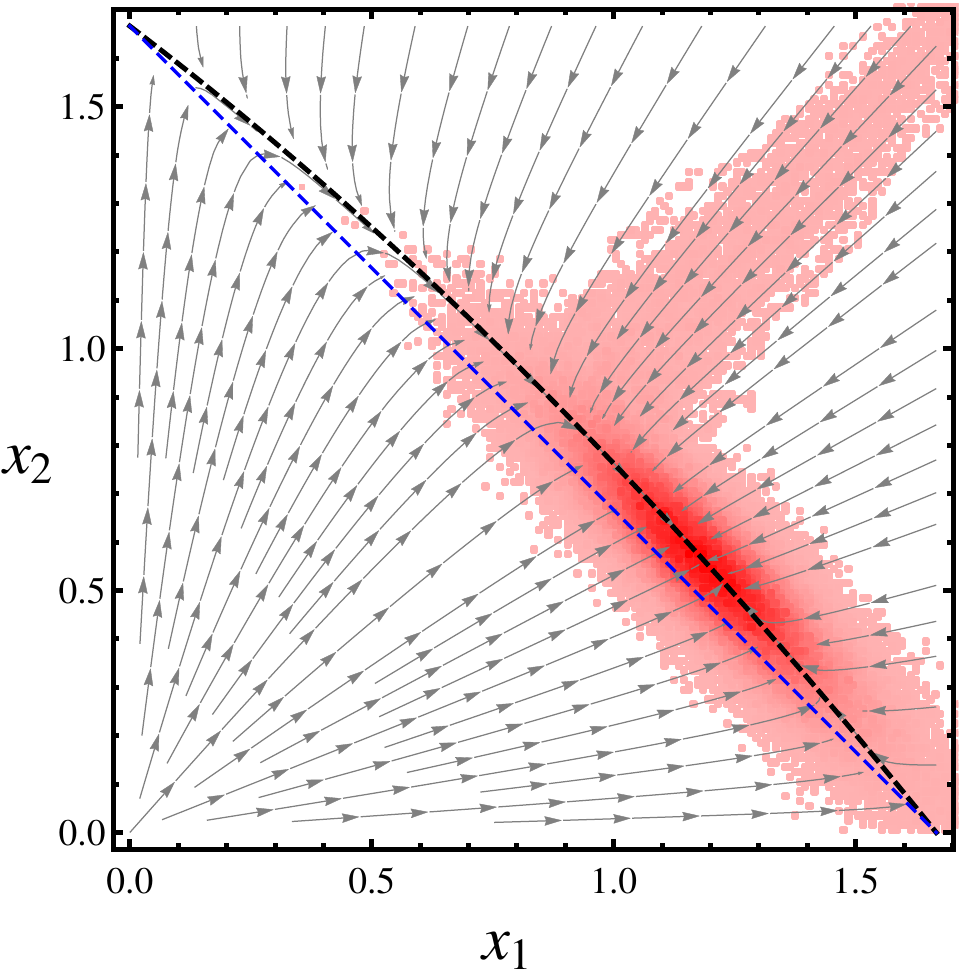}
\end{center}
\caption{(Color online) Deterministic trajectories of Eq.~\eqref{SDE_orig} for a non-neutral system. A histogram of stochastic trajectories is given in red. The black dashed line is the slow subspace Eq.\eqref{slow_subspace}. The blue dashed line is the CM from neutral theory. }\label{fig_traj_selection}
\end{figure}

The stochastic dynamics (shown in red in Fig.~\ref{fig_traj_neutral}) are dominated by the deterministic dynamics far from the CM, and there is a rapid collapse to its vicinity. Fluctuations taking the system too far away from the CM are similarly countered by the deterministic dynamics dragging the system back. The net result is a drift along the CM until either of the axes are reached and fixation of one of the types is achieved. This effect has also been noted and exploited in \cite{lin_mig_1,lin_mig_2} under the investigation of the evolution of dispersion. To encapsulate this behavior in a mathematical form we apply a methodology which we have recently used to reduce stochastic metapopulation models to effective models involving only one island~\cite{constable_phys,constable_bio}. The essential idea is to restrict the system to the CM, and to obtain the effective stochastic dynamics along the manifold by applying the projection operator 
\begin{equation}
P_{ij} = v^{(1)}_i u^{(1)}_j\label{proj_op}
\end{equation}
to the SDEs \eqref{SDE_orig} to eliminate the fast mode, keeping the slow mode intact. 

To carry out this program, we first rescale the $x_i$ and time in order to eliminate various constants from the calculation. Writing $y_{1} = c_0 x_1/(b_0 - d_0), y_{2}= c_0 x_2/(b_0 - d_0)$ and $\tilde{\tau}= (b_0 - d_0)\tau$, the equation of the CM becomes $y_{2}=1-y_{1}$ and the non-zero eigenvalue of the Jacobian is now equal to $-1$. Applying the condition $y_{2}=1-y_{1}$, gives $\bm{A} = 0$, confirming that there is no deterministic dynamics along the CM. We denote the coordinate along the CM as $z$, and choose this to be equal to $y_{1}$, although many other choices are possible. Since $\dot{y}_{2} =-\dot{y}_{1}$ on $y_{2}=1-y_{1}$, application of the projection operator to the left-hand side of Eq.~\eqref{SDE_orig}, and to the noise term on the right-hand side gives $\dot{z} = \zeta(\tilde{\tau})/\sqrt{V}$, where $\zeta(\tilde{\tau})=P_{11}\eta_{1}(\tilde{\tau})+P_{12}\eta_{2}(\tilde{\tau})$. It should be noted that since the projection operator depends on $y_{1}$, the direction of the dominant noise component changes, as can be seen from Fig.~\ref{fig_traj_neutral}. From the properties of $\eta_i$, we see that the effective noise $\zeta$ is Gaussian with zero mean and with correlator 
\begin{equation}
\left\langle \zeta(\tilde{\tau}) \zeta(\tilde{\tau}') \right\rangle = \left[ P^2_{11} B_{11}(\bm{y}) + P^2_{12} B_{22}(\bm{y}) \right] \delta\left( \tilde{\tau} - \tilde{\tau}' \right),
\label{zeta_correlator}
\end{equation}
with the $B_{ij}$ being evaluated on $y_{2}=1-y_{1}$. A calculation of the term in square brackets in Eq.~\eqref{zeta_correlator}, allows us to arrive at the following form for the SDE describing the dynamics after the fast-mode elimination:
\begin{equation}
\frac{\mathrm{d}z}{\mathrm{d}\tilde{\tau}} = \bar{A}(z) + \frac{1}{\sqrt{V}} \zeta(\tilde{\tau}),
\label{SDE_reduced}
\end{equation}
where $\bar{A}(z)=0$ and where $\zeta(\tilde{\tau})$ is a Gaussian noise with zero 
mean and correlator
\begin{equation}
\left\langle \zeta(\tilde{\tau}) \zeta(\tilde{\tau}') \right\rangle = \bar{B}(z) \delta\left( \tilde{\tau} - \tilde{\tau}' \right); \ \ 
\bar{B}(z) = 2 \frac{b_0 c_0}{(b_0 - d_ 0 )^2} z(1-z).
\label{correlator_reduced}
\end{equation}
If we define $N = (b_0 - d_0)V/c_{0}$ (the size of the population on the CM, Eq.~\eqref{centre_manifold}) then Eqs.~\eqref{SDE_reduced} and \eqref{correlator_reduced} are exactly the Moran model in rescaled time $\bar{\tau} = \left[b_{0}/(b_{0}-d_{0})\right]\tilde{\tau}$, where $z$ is the fraction of type $A_{1}$ alleles and $N$ the total population size~\cite{blythe_mckane_models_2007}. We therefore conclude that the neutral form of the SLVC model reduces to precisely the Moran model, under the fast-mode elimination procedure described in~\cite{constable_bio,constable_phys}.
\begin{figure}
\setlength{\abovecaptionskip}{-2pt plus 3pt minus 2pt}
\begin{center}
\includegraphics[width=0.4\textwidth]{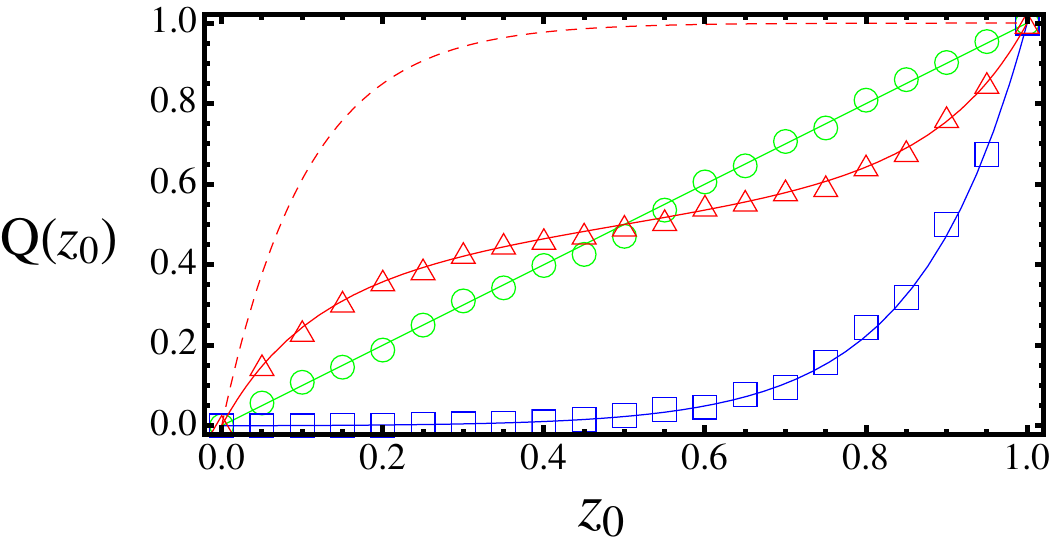}
\includegraphics[width=0.4\textwidth]{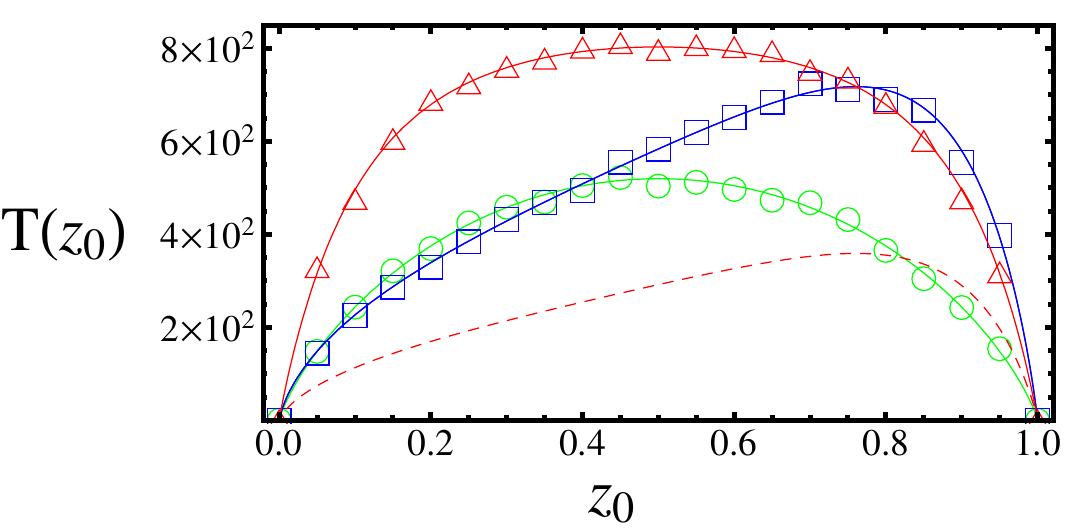}
\end{center}
\caption{(Color online) Probability of fixation, $Q(z_{0})$ and mean unconditional time to fixation $T(z_{0})$, where $z_0$ is the initial value of $z$ on the CM. Continuous lines are obtained from reduced theory and markers from Gillespie simulation. Green circles are obtained from a neutral system with parameters $V=150$, $b_{0}=3.1$, $d_{0}=1.1$ and $c_{0}=0.4$. Blue square markers with parameters $V=300$, $\epsilon=0.01$, $b_{0}=2$, $d_{0} = 1$, $c_{0}=0.2$, $\gamma_{11}=1$, $\gamma_{12}=2$, $\gamma_{21}=0$, $\Gamma=0$. Red triangles with $V=500$, $\epsilon= 0.015$, $b_{0}= 2$, $d_{0}=1$, $c_{0}=0.8$, $\gamma_{11}= \gamma_{22}=1$, $\gamma_{12}=\gamma_{21}=-1$. The red dashed line is the prediction of the standard Moran model with the same parameters as the SLVC model with the red solid line.}\label{Q_T_fig}
\end{figure}

It is now natural to ask what model is obtained by the elimination of the fast modes of the non-neutral SLVC model. As usual in population genetics, we work to linear order in the selection strength, and so begin by writing
\begin{equation}
b_i = b_0 \left( 1 + \epsilon \beta_i \right), \ 
d_i = d_0 \left( 1 + \epsilon \delta_i \right), \ 
c_{ij} = c_0 \left( 1 + \epsilon \gamma_{ij} \right), 
\label{epsilon_intro}
\end{equation} 
where $\epsilon$ is a small parameter which will later be related to the selection strength in the Moran model. The constants $\beta_i, \delta_i$ and $\gamma_{ij}$ are assumed to be of order one. Although for $\epsilon \neq 0$, there will not be a CM, we still expect there to be separation of timescales which will allow us to identify fast and slow variables. We pick out the slow subspace, and so eliminate the fast deterministic dynamics, by setting the product $\bm{u}^{(2)}\cdot\bm{A}(\bm{x})$ equal to zero~\cite{constable_phys}. This leads to an equation of the form $y_{2} = 1 - y_{1} + \epsilon f(y_{1}) + \mathcal{O}(\epsilon^2)$, where $f(y_{1})$ is quadratic in $y_{1}$. In order to make a comparison to the Moran model, we ask that this line passes through the points $\bm{y}=(1,0)$ and $\bm{y}=(0,1)$, which implies that $f(0)=0$ and $f(1)=0$, which leads to the two conditions 
\begin{equation}
\beta_i = \frac{\gamma_{ii}(b_0 - d_0) + d_0\delta_i}{b_0}, \ \ i=1,2.
\label{end_conditions}
\end{equation}
With this choice of the birth rates, the slow subspace takes the form
\begin{equation}
y_{2} = \left( 1 - y_{1} \right) \left[ 1 + \epsilon\,\Gamma\,y_{1} + \mathcal{O}\left( \epsilon^2 \right) \right],
\label{slow_subspace}
\end{equation}
where $\Gamma \equiv \gamma_{11} + \gamma_{22} - \gamma_{12} - \gamma_{21}$. It is interesting to note that the conditions in Eq.~\eqref{end_conditions} have also eliminated any reference to the death rates $\delta_i$, and that as long as the competition rates do not satisfy $\Gamma = 0$, the slow subspace will be curved. Simulations show that to an excellent approximation, the deterministic system collapses down to a line given by Eq.~\eqref{slow_subspace}, as shown in Fig.~\ref{fig_traj_selection}. The agreement persists even if we do not impose the conditions \eqref{end_conditions}, so that the line does not pass directly through the points $\bm{y}=(1,0)$ and $\bm{y}=(0,1)$.

The effective stochastic dynamics on the slow subspace is found by applying the same arguments as in the neutral case. A key aspect of the approximation is that the same form of the projection operator will be used when $\epsilon \neq 0$ as was used when $\epsilon = 0$. In previous applications~\cite{constable_phys,constable_bio} this was found to be a very good approximation, and we will see that a similar conclusion applies in the current case. Therefore applying $P_{ij}$ given by Eq.~\eqref{proj_op} to Eq.~\eqref{SDE_orig} gives Eq.~\eqref{SDE_reduced}, but now with
\begin{equation}
\bar{A}(z) = \epsilon z \left( 1 - z \right) \left[ \left( \gamma_{11} - \gamma_{12} \right)  - \Gamma z + \mathcal{O}\left( \epsilon \right) \right].
\label{A_bar}
\end{equation}

To test the validity of the fast-mode elimination procedure we compare the results for the probability of fixation and the time to fixation found from the reduced model to Gillespie simulations of the original IBM~\cite{gillespie_1977}. Both of these quantities can be found, either analytically or numerically, from the backward FPE corresponding to the reduced SDE \eqref{SDE_reduced}~\cite{risken_1989}. Fig.~\ref{Q_T_fig} shows the reduced model captures the properties of the full model extremely well.

The form of the reduced drift coefficient in the SLVC model given by Eq.~\eqref{A_bar} can give very different results to that of the Moran model, which has $A(z)=sz(1-z)$, for selection strength $s$. Only if $\Gamma = 0$ is the reduced SLVC (in terms of the time variable $\bar{\tau}$) equivalent to the Moran model, with selection strength $s=(b_0 - d_0)(\gamma_{11}-\gamma_{12})\epsilon/b_0$. If $\Gamma \neq 0$, the deterministic dynamics will have a fixed point at $z^* = \phi_1/\Gamma$, where $\phi_1 \equiv  \gamma_{11}-\gamma_{12}$. This fixed point is in the range $0 < z < 1$ only if (i) $\phi_1, \phi_2 > 0$, or (ii) $\phi_1, \phi_2 < 0$, where $\phi_2 \equiv  \gamma_{22} -\gamma_{21}$. In case (i) the fixed point is stable, in case (ii) it is unstable. On general biological grounds, we would expect intraspecific competition to be stronger than interspecific competition \cite{connell_1983}, which would imply that both $\phi_1$ and $\phi_2$ are positive, and so point to the existence of a stable fixed point of the reduced SLVC model dynamics. In Fig.~\ref{Q_T_fig} we show how the existence of such a stable fixed point ensures that the reduced SLVC model gives qualitatively different results to that of the Moran model with the same value of $s$. 

In this Letter we have taken a more ecologically motivated view of genetic drift, by basing it on the SLVC model, instead of starting from a fixed-size population model, such as the Moran model. By applying a well-defined, and remarkably accurate, approximation scheme to the SLVC, we showed that the reduced model could in general only be identified as a Moran model if the theory was neutral. If selection was present, the resulting model had additional features not present in the simple Moran model, such as the possibility of a fixed point away from the boundaries. We expect this to remain true in more complex situations, and hope to report on this elsewhere.

\begin{acknowledgments}
G.W.A.C. thanks the Faculty of Engineering and Physical Sciences, University of Manchester for funding through a Dean's Scholarship.
\end{acknowledgments}


\end{document}